\documentclass{PoS}

\usepackage{graphics,mathbbol}

\usepackage{epsfig,multicol,multirow}
\usepackage{amsmath,subfigure,latexsym,amssymb}

\def\be{\begin{equation}}               
\def\ee{\end{equation}}           

\def\bea{\begin{eqnarray}}               
\def\eea{\end{eqnarray}}    

\def\lsi{\lesssim}
\def\gsi{\gtrsim}

\newcommand{\nn}{\nonumber}

\newcommand{\beq}{\begin{equation}}
\newcommand{\eeq}{\end{equation}}
\newcommand{\uno}{1 \!\!\! 1}

\def\lsi{\raise0.3ex\hbox{$<$\kern-0.75em\raise-1.1ex\hbox{$\sim$}}}
\def\gsi{\raise0.3ex\hbox{$>$\kern-0.75em\raise-1.1ex\hbox{$\sim$}}}

\def\backder{\raise1.4ex\hbox{$\leftarrow$\kern-0.75em\raise-1.4ex\hbox{$\partial$}}}

\def\forder{\raise1.4ex\hbox{$\rightarrow$\kern-0.75em\raise-1.4ex\hbox{$\partial$}}}

\newcommand{\NN}{{\kern+.25em\sf{N}\kern-.78em\sf{I} \kern+.78em\kern-.25em}}

\title{Nucleon structure in terms of OPE with non-perturbative Wilson coefficients}

\ShortTitle{Nucleon structure in terms of OPE with non-perturbative Wilson coefficients}

\author{\speaker{W.\ Bietenholz}$^{\rm \ a,b}$, N.\ Cundy$^{\rm \ b}$, M.\ G\"{o}ckeler$^{\rm \ b}$, R.\ Horsley$^{\rm \ c}$, \newline H.\ Perlt$^{\rm \ d}$, D.\ Pleiter$^{\rm \ a}$, P.E.L.\ Rakow$^{\rm \ e}$, G.\ Schierholz$^{\rm \ a}$, \newline A.\ Schiller$^{\rm \ d}$ and J.M.\ Zanotti$^{\rm \ c}$%
        \\
\ \\
$^{\rm \ a}$ John von Neumann Institut f\"{u}r Computing NIC, \\
~~~~Deutsches Elektron-Synchrotron DESY, 15738 Zeuthen, Germany \\
$^{\rm \ b}$ Institut f\"{u}r Theoretische Physik, 
Universit\"{a}t Regensburg, 93040 Regensburg, Germany \\
$^{\rm \ c}$ School of Physics, University of Edinburgh, Edinburgh EH9 3JZ, United Kingdom \\
$^{\rm \ d}$ Institut f\"{u}r Theoretische Physik, 
Universit\"{a}t Leipzig, 04109 Leipzig, Germany \\
$^{\rm \ e}$ Theoretical Physics Division, Dept.\ of Mathematical Sciences, University of Liverpool, \\ ~~~~Liverpool, L69 3BX, United Kingdom \\
\ \\
E-mail: \email{bietenho@ifh.de} \\ }

\abstract{Lattice calculations could boost our understanding
of Deep Inelastic Scattering by evaluating moments of the Nucleon 
Structure Functions. To this end we study the product of
electromagnetic currents between quark states. The
Operator Product Expansion (OPE) decomposes it into matrix elements 
of local operators (depending on the quark momenta) and Wilson coefficients 
(as functions of the larger photon momenta). For consistency 
with the matrix elements, we evaluate 
a set of Wilson coefficients non-perturbatively, based on 
propagators for numerous momentum sources, on a $\, 24^3 \times 48\,$ 
lattice. The use of overlap quarks suppresses unwanted operator mixing and
lattice artifacts. Results for the leading Wilson coefficients 
are extracted by means of Singular Value Decomposition.}

\FullConference{The XXVI International Symposium on Lattice Field Theory \\
                 July 14 - 19, 2008\\
                 Williamsburg, Virginia, USA}

\begin{document}

\section{Motivation}

The computation of moments of the {\em nucleon structure functions}
is a fascinating challenge: it is known to be difficult, but
it is a point where lattice results could
contribute much to the understanding and interpretation
of phenomenological data from Deep Inelastic Scattering.

The continuum formulation of hadron structure functions
is plagued by renormalon ambiguities, {\it i.e.}\ power-like
IR contributions, see {\it e.g.}\ Refs.\ \cite{renormalon}.
Here we refer to the lattice regularisation, where
a general moment of the nucleon structure function can be expanded as
\be
{\cal M}(q^{2}) = c^{(2)} (aq) A_{2} (a) +
c^{(4)} (aq) \frac{1}{q^{2}} A_{4}(a) + \dots \ 
\{ \ {\rm higher~twists} \ \} \ .
\ee
It depends on the transfer momentum $q$,
while $a$ is the lattice spacing, $c^{(2)}$, $c^{(4)} \dots$ 
are Wilson coefficients (where the superscript is the twist),
and $A_{2}$, $A_{4} \dots$ are reduced matrix elements 
(``reduced'' in the sense that the Lorentz structure is factored out). 

For the evaluation of $A_{2}$ there is an
established procedure, which employs the ratio between two-point
and three-point correlation functions \cite{A2rat}. On the other hand,
the Wilson coefficients have usually been evaluated in continuum
perturbation theory. However, we need a cancellation of
singularities in the terms $c^{(2)} \, , \ A_{4} \propto 1/a^{2}$,
which are again a facet of the renormalon problem. This
requires a strictly consistent treatment \cite{MaSa}. Therefore 
we evaluate the Wilson coefficients non-perturbatively as well.
This method is particularly suitable for disentangling higher
twist contributions.
The use of overlap quarks suppresses undesired operator mixings.

In this report we present new numerical results for $c^{(2)}$.
For earlier results with Wilson fermions we refer to Ref.\ 
\cite{Wilfer}. In a previous study with overlap quarks on a 
$16^{3} \times 32$ lattice \cite{Lat07} some problems persisted,
which motivated us to enlarge the lattice to the size 
$24^{3} \times 48$.

\section{Operator Product Expansion on the lattice}

To be explicit, we apply the Operator Product Expansion (OPE)
to a product of electromagnetic currents
$J_{\mu}$ between quark states $ | \psi (p) \rangle$,
\bea
W_{\mu \nu} (p,q) &=& \langle \psi (p) | J_{\mu}(q) 
J_{\nu}^{\dagger}(q) | \psi (p) \rangle \nn \\
&=& \sum_{m} C_{\mu \nu,i,\mu_{1}\dots \mu_{n}}^{(m)} (q) \
\langle \psi (p) | {\cal O}^{(m)}_{i, \mu_{1} \dots \mu_{n}}
| \psi (p) \rangle \ .  \label{OPEeq}
\eea
The lower line is a decomposition in terms of {\em local}
operators ${\cal O}^{(m)}$, which characterise the nucleon 
structure. The index $i= 1 \dots 16$ specifies the
Dirac structure, $\mu_{1} \dots \mu_{n}$ indicate the
momentum components involved, and $m$ distinguishes operators with 
the same symmetry. Note that the corresponding Wilson coefficients
$C^{(m)}$ depend solely on the transfer momentum $q$.

A truncation of this expansion at some low operator dimension
requires the following {\em scale separation,}
\be  \label{sepa}
p^{2} \ll q^{2} \ll \Big( \frac{\pi}{a} \Big)^{2} \ .
\ee
So far we have been using periodic boundary conditions for the gauge 
field. Thus a sizable lattice is required to have a 
set of small momenta $p^{2}$ available.
If this scale separation is realised, it justifies a truncation
of the OPE (\ref{OPEeq}), which cuts off operators with high
derivatives, {\it i.e.} high powers of $p$. Here we consider 
quark bilinears with up to 3 derivatives,
\vspace*{-0.6mm}
\be
\bar \psi \, \Gamma_{i} \psi \ , \quad
\bar \psi \, \Gamma_{i} D_{\mu_{1}} \psi \ , \quad
\bar \psi \, \Gamma_{i} D_{\mu_{1}} D_{\mu_{2}} \psi \ , \quad
\bar \psi \, \Gamma_{i} D_{\mu_{1}} D_{\mu_{2}} D_{\mu_{3}} \psi \ ,
\ee
\vspace*{-0.6mm}
where $\Gamma_{i}$ runs over a basis of the
Clifford algebra. This still amounts to an apparently frightening
set of $\, 16 \cdot \sum_{d=0}^{3} 4^{d} = 1360 \, $
operators. However, we choose the {\em isotropic} transfer momenta
\vspace*{-1mm}
\bea
q_{a} &=& \frac{\pi}{4a} (1,1,1,1) \qquad
(|q_{a}| \simeq 4.1~{\rm GeV}) \ , \nn \\
q_{b} &=& \frac{\pi}{3a} (1,1,1,1) \qquad
(|q_{b}| \simeq 5.5~{\rm GeV}) \ .  \label{isomom}
\eea
In this case the symmetry reduces the set of operators
to only 67 equivalence classes \cite{Lat07}. We denote them
as $C_{1} \dots C_{67}$, which refer to an ascending number 
of derivatives:
\begin{eqnarray*}
C_{1} & :& {\rm no~derivative,~vanishes~in~the~chiral~limit,~
multiplies~} \bar \psi \uno \psi \\
C_{2} \dots C_{6} &:& {\rm one~derivative, Bjorken~scaling} \propto 1/q^{2} \\
C_{7} \dots C_{16} &:& {\rm two~derivatives,~vanish~in~the~chiral~limit} \\
C_{17} \dots C_{67} &:& {\rm three~derivatives,~Bjorken~scaling}
\propto 1/(q^{2})^{2} \ . 
\end{eqnarray*}
The coefficients of terms with an even number of derivatives
vanish at quark mass $m=0$ due to chiral symmetry. In the Bjorken
limit of large $q^{2}$ the coefficients of terms with one (three)
derivative(s) are expected to scale as $\, C_{m} \propto 1/q^{2}$ \ 
($C_{m} \propto 1/(q^{2})^{2})\,$.

To separate the scales even better, we have now implemented
twisted boundary conditions: thus very small $p^{2}$ become
accessible, which enables us to use $q^{2}$ further below
the momentum cutoff squared \cite{prep}. 
This allows us to consider for instance $q = \frac{\pi}{6a} (1,1,1,1)$.

\section{Results for the Wilson coefficients}

In our numerical study we analysed configurations that were generated 
in the quenched approximation 
with the L\"{u}scher-Weisz gauge action in a volume $V = 24^{3} \times 48$,
with a physical lattice spacing $a \simeq 0.075~{\rm fm}$.
To provide finite OPE matrix elements, as they occur in eq.\ (\ref{OPEeq}), 
we fixed the lattice Landau gauge; this is also favourable to reduce
the statistical noise.

For the valence quarks we used overlap fermions with the
parameter $\rho = 1.4$ (negative mass of the Wilson kernel)
at a bare quark mass of $0.028$ in lattice units, which corresponds
to about $73~{\rm MeV}$. The application of chiral lattice fermions
suppresses undesired operator mixings, as well as $O(a)$ lattice 
artifacts.

We denote the number of $p$-sources as $M$, and the number of
Wilson coefficients to be considered as $C_{1} \dots C_{N}$,
with $N=67$ in our case.
So far we evaluated $W_{\mu \nu}$ off-shell for $M=25$ $p$-momentum 
sources at transfer momentum $q_{a}$ (on different configurations), 
and for $M=10$ $p$-momentum sources at $q_{b}$. 
In each case this yields a system constraining the Wilson coefficients,
\be  \label{megaeq}
\left( \begin{array}{c} 
W^{(p_{1})} \\ . \\ . \\ . \\  W^{(p_{M})}
\end{array} \right) = \left( \begin{array}{ccccc}
{\cal O}_{1}^{(p_{1})} & . & . & . & {\cal O}_{N}^{(p_{1})} \\
 . &  . & . & . & . \\
 . &  . & . & . & . \\
 . &  . & . & . & . \\
{\cal O}_{1}^{(p_{M})} & . & . & . & {\cal O}_{N}^{(p_{M})} 
\end{array} \right) \
\left( \begin{array}{c} 
C_{1} \\ . \\ . \\ C_{N} \end{array} \right) 
\ee
where the elements $W^{(p_{i})}$ and ${\cal O}_{k}^{(p_{i})}$
are $4 \times 4$ matrices.

\begin{figure}[h!]
\begin{center}
\includegraphics[angle=270,width=.7\linewidth]{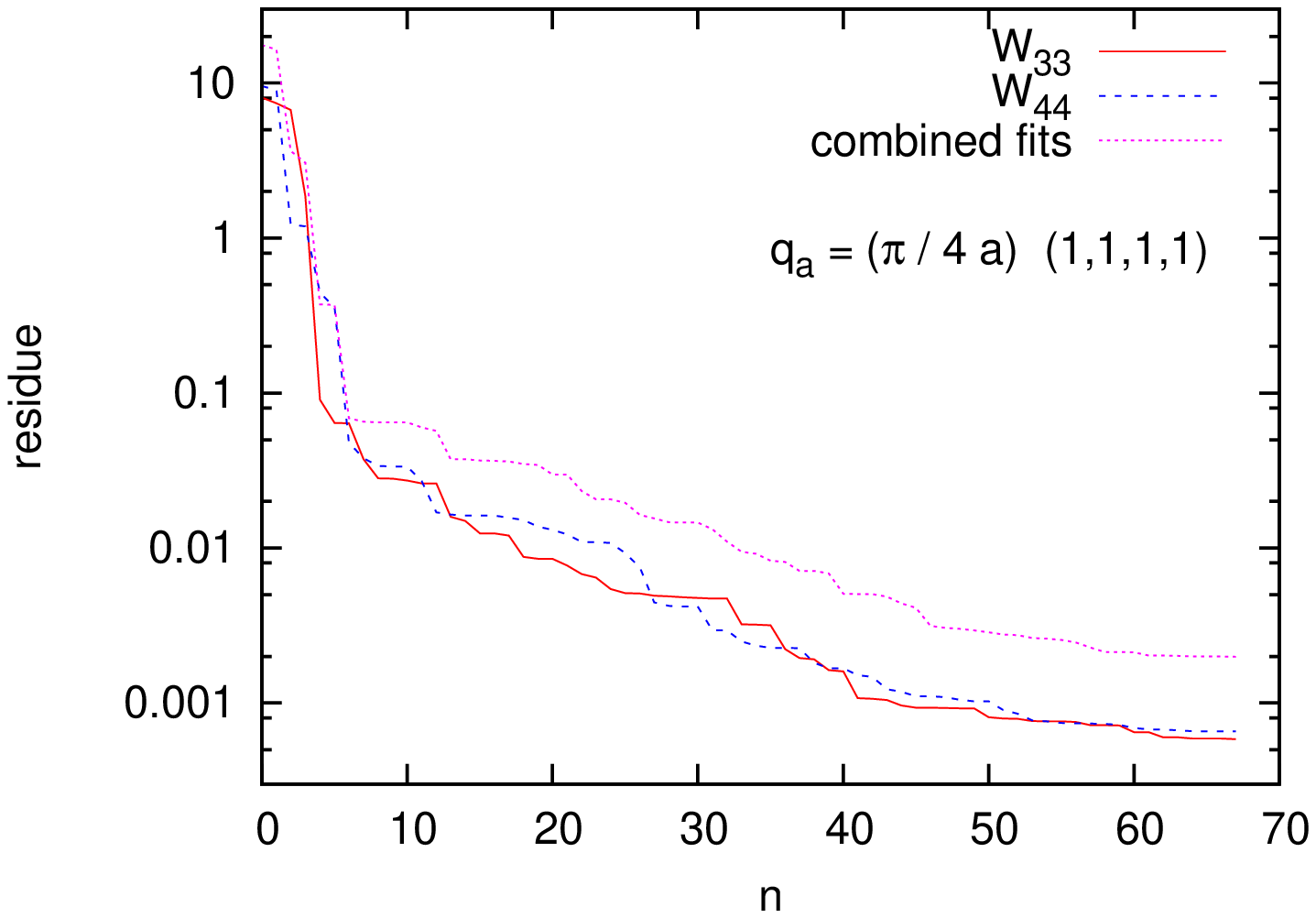}
\includegraphics[angle=270,width=.7\linewidth]{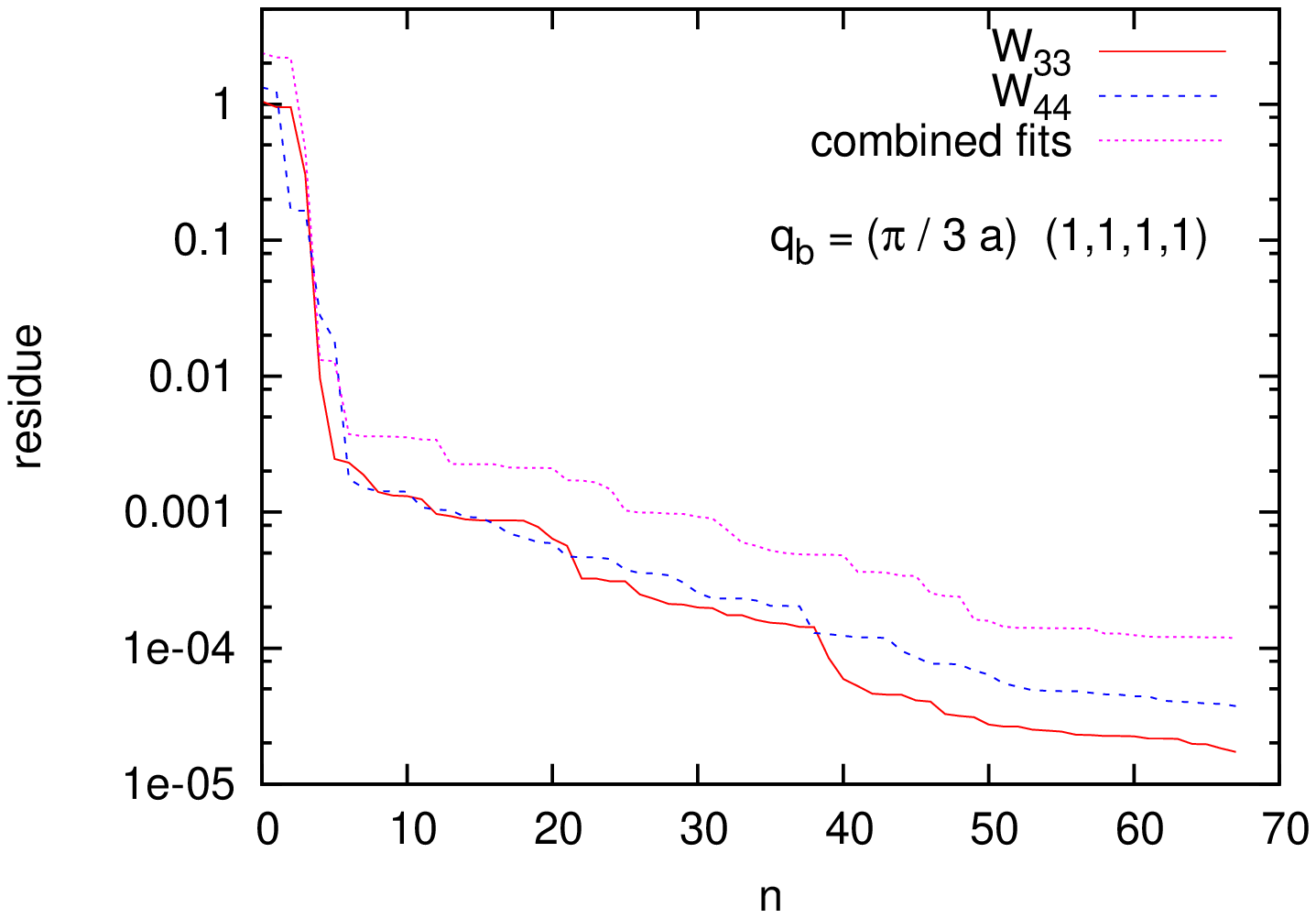}
\end{center}
\caption{The convergence of the Wilson coefficients as the 
number of the $n$ most powerful constraints is increased. We show the
residues as a function of $n$ for the transfer momentum 
$q_{a}$ with 25 $p$-sources (upper plot), and for $q_{b}$ with 10 
$p$-sources (lower plot). In both cases we see a convincing
convergence before $n$ reaches the deterministic number $67$, which
confirms that the extracted Wilson coefficients are trustworthy.}
\label{converge}
\end{figure}
In both cases the system is over-determined since $16 M > N$. 
We apply {\em Singular Value Decomposition} as an established method to 
analyse such systems \cite{NumRep}. Roughly speaking,
this methods selects the $n \leq N$ conditions with the
``maximal impact'' on the solution $C_{1} , \dots , C_{N}$.
A rapid convergence for increasing $n$ approves a reliable 
result ({\it i.e.}\ the remaining conditions are negligible).
With 12 singular values (analogues of eigenvalues in rectangular
matrices \cite{NumRep}) we do observe this feature. As an example
Figure \ref{converge} shows the convergence in $n$ for the 
Wilson coefficients under consideration, which saturates 
around $n \approx 50$ for $q_{a}$ and for $q_{b}$. 
In the latter case, the total number of constraints is lower, 
hence it is easier to satisfy them to a good precision.
Therefore the residues\footnote{The residue is the norm of the
difference between the two sides of eq.\ (\ref{megaeq}) in the
given approximation.} are smaller, but the result is less reliable. 
Note also that $q_{a}^{2} = \frac{1}{4}
(\pi /a)^{2}$ is more promising than $q_{b}^{2} = \frac{4}{9}
(\pi /a)^{2}$ in view of the required scale separation (\ref{sepa}).

\begin{figure}[h!]
\begin{center}
\includegraphics[angle=270,width=.8\linewidth]{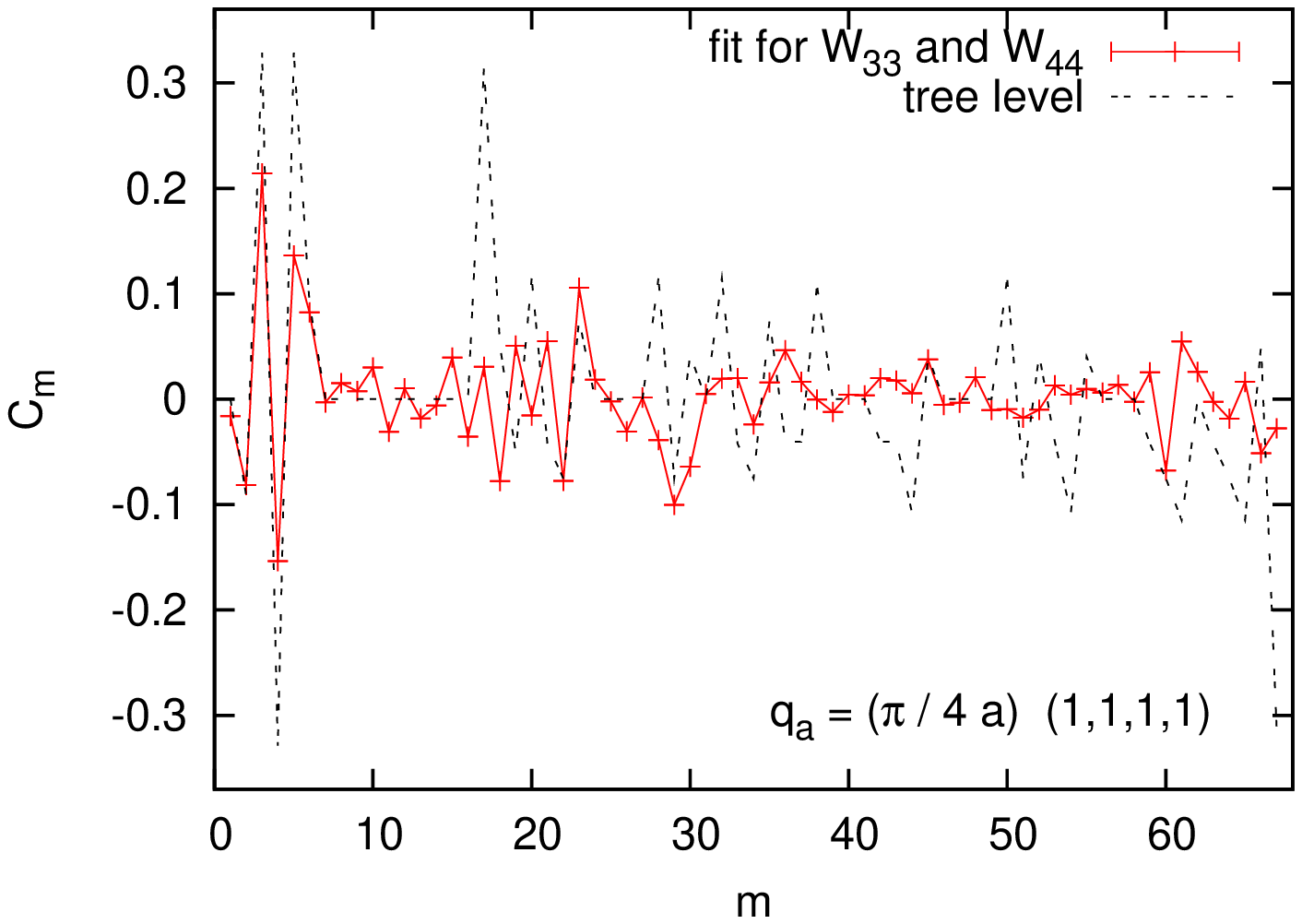}
\includegraphics[angle=270,width=.8\linewidth]{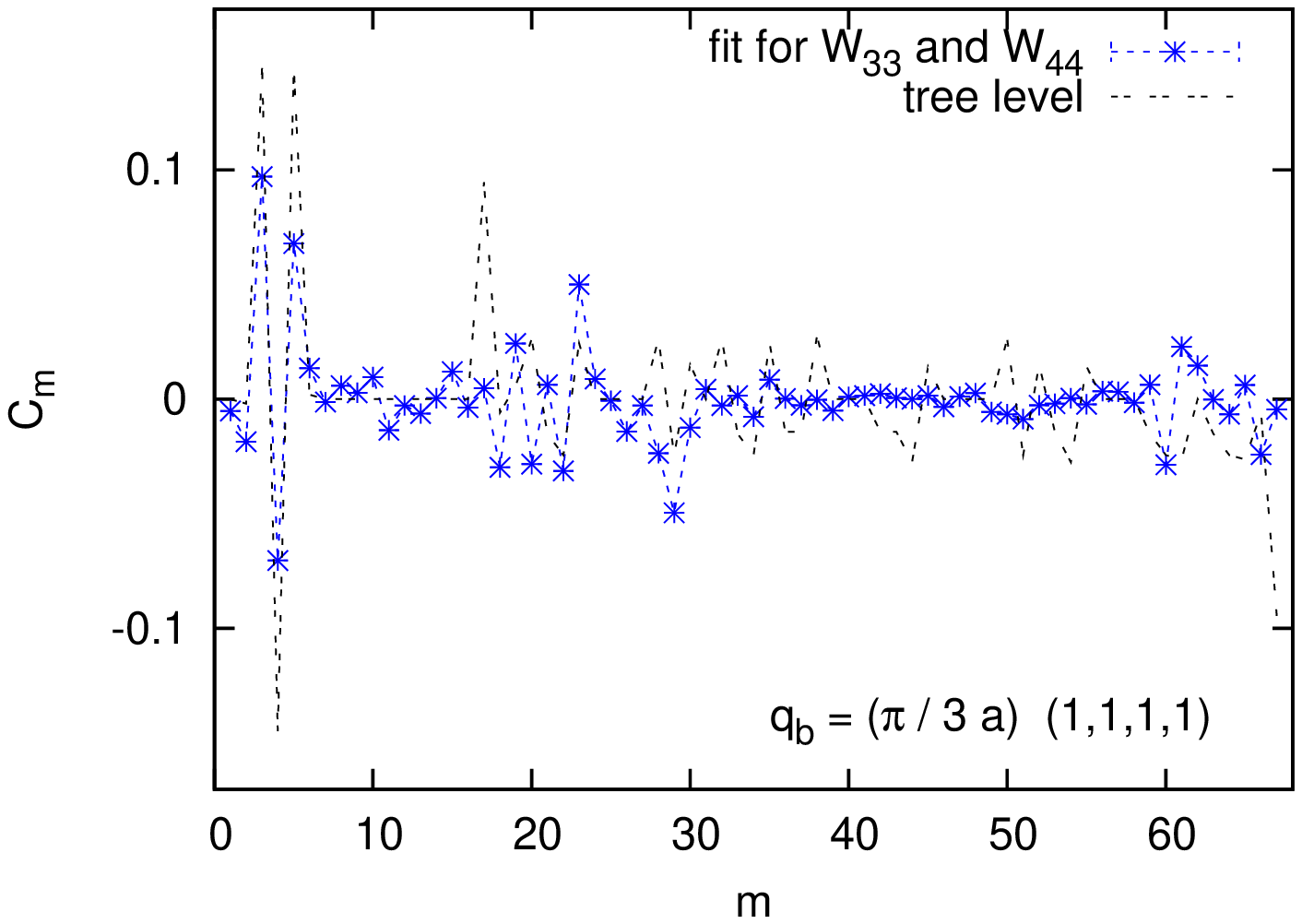}
\end{center}
\caption{The Wilson coefficients evaluated at momentum transfer
$q_{a}$ (upper plot) and $q_{b}$ (lower plot). The results are
compared to the corresponding tree level values 
(with the same overlap Dirac operator). We observe the same
pattern, but a significant non-perturbative correction. 
In the chiral limit $C_{1}$ and $C_{7} \dots C_{16}$ vanish. 
The small values that we obtain (with light quarks)
for these coefficients provide a consistent picture.}
\label{Cm}
\end{figure}

There are a number of discrete rotations and reflections which
leave our isotropic $q$-momenta (\ref{isomom}) invariant. For example 
this allows us to exchange $W_{33}$ with $W_{44}$. As a consequence,
specific pairs of Wilson coefficients belonging to different
operators have to coincide; for instance the coefficient
of ${\cal O}_{i,33}$ in $W_{33}$ is equal to the coefficient
of ${\cal O}_{i,44}$ in $W_{44}$.
We implemented this property in combined fits, which have of 
course somewhat larger residues, see Figure \ref{converge}.\\

We now proceed to the actual results for the Wilson coefficients
obtained from $W_{33}$ and $W_{44}$. The best results emerge 
from combined fits. They are shown in Figure \ref{Cm}, 
which also displays the corresponding tree level values for comparison.

In the limit of zero quark mass the coefficients $C_{1}$ and
$C_{7} \dots C_{16}$ vanish due to chiral symmetry, as we 
anticipated in Section 2. Since we are dealing with
light quarks represented by chiral lattice fermions,
we are fairly close to chirality. Therefore it
is a stringent consistency test that we do obtain
particularly small values for these coefficients. 
That property had not been observed with Wilson fermions 
\cite{Wilfer}.

\begin{figure}[h!]
\begin{center}
\includegraphics[angle=270,width=.8\linewidth]{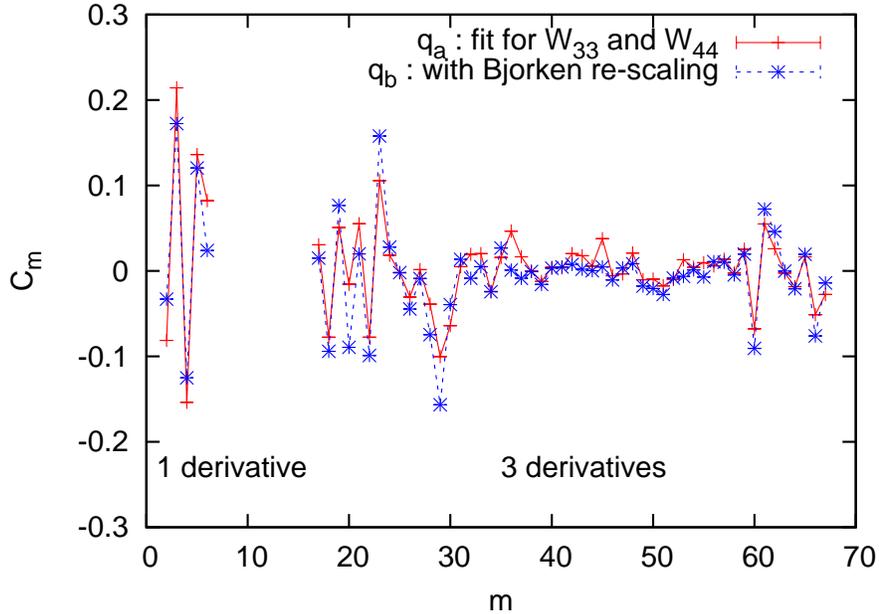}
\end{center}
\caption{The Wilson coefficients evaluated at momentum transfer
$q_{a}$ and $q_{b}$, where in the latter case the coefficients
of terms with one or three derivatives are amplified according
to the expected scaling in the Bjorken limit. In fact the
$C_{m}$ are now in the same range, and we observe a very
satisfactory level of agreement.}
\label{Bjo}
\end{figure}

Finally we also test the Bjorken scaling property, which we
mentioned in Section 2. We show again the $C_{m}$ obtained at
$q_{a}$. We compare them to the Wilson coefficients evaluated 
at $q_{b}$, where $C_{2} \dots C_{6}$ are enhanced with a factor 
$q_{b}^{2} / q_{a}^{2} = 16/9$,
and $C_{17} \dots C_{67}$ are amplified with the square of this
factor. The Bjorken re-scaled results are confronted in Figure
\ref{Bjo}, and we see an impressive similarity, in
particular for cases with relatively large $|C_{m}|\,$.

\section{Conclusions and outlook}

Our method to evaluate the Wilson coefficients non-perturbatively
works successfully. We arrived at conclusive non-perturbative
results for a considerable set of Wilson coefficients. We are 
continuing to evaluate constraints for additional $q$ and $p$ 
momenta to further tighten the control over possible artifacts.
The use of twisted boundary conditions enables us now to achieve
an even better scale separation.\\

In a final step, the nucleon structure function
${\cal M}$ is obtained by {\em Nachtmann integration} \cite{Nacht} over
$W_{\mu \nu}$. For instance, for the second moment this integral takes 
the form
\begin{eqnarray}
{\cal M}_{2} (q) &=& \frac{3 q^{2}}{ (4 \pi )^{2}} \int d \Omega_{q} \,
n_{\mu} \Big[ W_{\mu \nu}(q) - \frac{1}{4} \delta_{\mu \nu} W_{\rho \rho} \Big]
n_{\nu} \nn \\
& \to & \int_{0}^{1} dx \, \Big[ F_{2} (x,q^{2}) + 
\frac{1}{6} F_{L}(x,q^{2}) \Big] \ ,
\end{eqnarray}
where the Bjorken limit is taken in the lower line. The projection
vector has length $n^{2} =1$. 
For a different projection one obtains $\, \int_{0}^{1} 
dx \, [F_{2} - \frac{3}{2} F_{L}] \,$ instead.
Thus the combination of different projections
determines the {\em longitudinal structure function}
$ F_{L} = F_{2} - 2x F_{1} $. \\

With completed data, we will obtain a fully non-perturbative 
and consistent evaluation of the Nucleon Structure Functions \cite{prep}.\\

\noindent
{\bf Acknowledgement :} The computations
were performed on the clusters of the ``Norddeutscher
Verbund f\"{u}r Hoch- und H\"{o}chstleistungsrechnen'' (HLRN).
We thank Hinnerk St\"{u}ben for technical assistance.

\end{document}